\documentclass[12pt]{iopart}

\usepackage{graphicx}
\usepackage{tikz}
\usepackage[RPvoltages]{circuitikz}
 
\usepackage[english]{babel}
\usepackage{cite}
\usepackage[utf8]{inputenc}
\usepackage{lipsum}
\usepackage{url}
\usepackage{textcomp}
\usepackage{xcolor}
\usepackage{float}
\usepackage{blindtext}
\usepackage{subfigure}

\begin{document}

\title{Seeing the invisible: convection cells revealed with thermal imaging}

\date{today}

\author{Renzo Guido$^{1}$, Mateo Dutra$^{1}$, Martín Monteiro$^{1,2}$, Arturo C. Martí $^{1}$}
\address{$^1$Instituto de Física, Universidad de la República, Montevideo, Uruguay; $^2$Universidad ORT Uruguay, Montevideo, Uruguay}

\date{\today}

\ead{marti@fisica.edu.uy}

\begin{abstract}
Fluid instabilities are ubiquitous phenomena of great theoretical and applied importance. In particular, an intriguing example is the  thermocapillary or Bénard-Marangoni instability  which occurs when a thin  horizontal fluid layer, whose top surface is free, is heated from below.  In this phenomenon, after passing a certain temperature difference threshold, the fluid develops a regular pattern, usually hexagonal, of convection cells known as Bénard convection. In general this  pattern is not visible to the naked eye unless specific tracers are incorporated into the fluid. The use of thermal imaging  is a simple alternative not only for directly observing these phenomenon but also for obtaining  valuable quantitative information, such as the relationship between the critical wavelength and the thickness of the fluid layer. Here, we propose an  experiment specially suited for laboratory courses in fluid mechanics or nonlinear physics that involves the use of thermal cameras, or appropriate smartphone accessories, to study Bénard convection.
\end{abstract}


\section{Introduction}

In fluid dynamics, two distinct opposite behaviours can be distinguished (see for example \cite{cengel2013fluid}). On the one hand, when viscous effects predominate, for example in the case of a sphere moving slowly in a solution of water and glycerine, the flow is called \textit{laminar}. On the other hand, for example around a ball thrown at high speed in air, nonlinear terms dominate and the flow is known as \textit{turbulent}. In laminar flows it is often possible to solve the equations completely, whereas in turbulent flows it is generally not possible and a statistical approach must be adopted. The transition from laminar to turbulent flow when a control parameter is varied offers very important challenges from the theoretical and experimental point of view. In many cases a steady state loses its stability to give rise to more complex states, a phenomenon known as \textit{instability} that often leads to the formation of spatio-temporal patterns in the fluid \cite{cohen2004fluid,charru2011hydrodynamic,chandrasekhar2013hydrodynamic}. Being able to predict when a certain instability occurs or what are the characteristics of the patterns formed is of crucial importance in many applications.

Experimenting with, or simply observing, instabilities in fluids presents some peculiarities \cite{charru2011hydrodynamic}. The Kelvin Helmholtz instability that occurs when there is a difference in velocities at an interface can be observed in atmospheric phenomena but laboratory experimentation requires careful separation of fluid layers or strata. Taylor instability, which occurs between concentric cylinders rotating at different velocities, offers difficulties in terms of experimental apparatus and visualization. 
Another paradigmatic example is the thermal or Rayleigh-Bénard instability induced by density variations within a fluid due to temperature differences. The best known example concerns a fluid layer subjected to a vertical temperature gradient produced by two flat plates. When the lower plate is heated beyond a certain threshold, the initially stable quiescent state loses stability to give rise to the formation of convection cells.  Within this family of thermal instabilities, when we focus on a thin fluid layer whose upper surface is open to the atmosphere, surface tension effects also play a role. In this case we refer to 
thermocapillary or Bénard-Marangoni instability 

The Bénard-Marangoni instability which occurs when a free-surface fluid layer is heated from below is easy to implement experimentally but the visualisation and the extraction of quantitative data offers some difficulties. In this experiment when the temperature difference reaches a certain threshold, a hexagonal spatio-temporal pattern composed of convection cells is developed. This regimen is known as Bénard convection. Although this pattern is not visible to the naked eye, it can be directly observed using a thermal camera.

Thermal cameras are a valuable tool for experiments that require measuring temperature fields with require a certain spatial resolution. They enable the observation of phenomena that are not visible to the human eye. In university courses, they are particularly useful for conducting demonstrative classes and laboratory experiments  \cite{vollmer2001there,mollmann2007infrared,haglund2015thermal}. Furthermore, the cost of thermal cameras has decreased in recent years, resulting in their increased use in various physics topics and the proposal of numerous related activities \cite{xie2011infrared,oss2015electro,gfroerer2015thermal,vollmer2018infrared,kacovsky2019electric}. Their increasing affordability has made them more accessible to a wider range of users, including students and researchers, and they are increasingly being integrated into university courses to enhance the learning experience \cite{kacovsky2018thermal,planinvsivc2022infrared,Hong_2023}.  An alternative, even cheaper than conventional thermal cameras, are devices such as the Flir One, which are devices that attach to smartphones to convert them into a thermal camera. In this case, in addition to the facilities offered by thermal cameras, the capabilities of smartphones have proven to be very useful in a wide range of physics experiments in recent years \cite{monteiro2021using,monteiro2022resource}. In this article, we propose an experiment to study  important concepts in fluids, both qualitatively and   quantitative, related to  Bénard-Marangoni instability and thermal convection. 

\section{Bénard-Marangoni instability}

In 1901, H. Bénard experimented with thin fluid layers heated from below and found  the instability of heat conduction and the formation of cellular convection cells \cite{benard1901tourbillons}. Later, in 1916, Lord Rayleigh analysed theoretically the phenomenon and concluded that this instability was produced by the buoyancy force due to the differences in density between hot and cold fluid \cite{rayleigh1916lix}. Nevertheless, J. Pearson \cite{pearson1958convection}, proposed another mechanism as responsible for the convection. As in Bénard experiment the top surface is free to the atmosphere  the surface tension plays a significant role in the instability. Surface tension, as density, exhibits a strong dependence on the temperature, so he argues that in thin layers, the difference in surface tension was the driving mechanism of the instability. Years later, D. Nield  \cite{nield1964surface} studied both effects simultaneously and determined  that both effects contribute to the instability and quantified their effect. 

Bénard convection is characterized by hexagonal-shaped convection cells. When the temperature difference between the bottom surface and the surface reaches a certain threshold the quiescent ground state loses stability and convection cells appear \cite{cohen2004fluid}. The wavelength, or characteristic length, $\lambda$, is defined as the mean distance between the centres of the convection cells. Theoretical developments based on linear stability theory predict, at the threshold of instability, a linear relationship between the wavelength and the thickness of the fluid layer, $d$, as  
\begin{equation}
\lambda = 2.342 d, 
\label{eq_lambda}
\end{equation}
 valid  faraway from the borders of the container \cite{chandrasekhar2013hydrodynamic}.

Bénard-Marangoni instability is usually visualized with tracers, like graphite or aluminium  powder dispersed on the fluid, that in general affect to some extent the dynamics. In this way, it is possible to directly observe, with the naked eye, convection cells and in particular, measure their appearance, distribution and  size. To measure the  critical temperature another approach must be adopted such as the use of thermocouples. In this case it is necessary to introduce two thermocouples one touching the bottom surface and the other on the top (free) surface. 
Due to the non-negligible size of the thermocouples compared to the cell,
usually a few millimeters, this procedure involves some difficulties, in particular affects  boundary conditions on the fluid layer. Nevertheless, with this procedure  the temperature difference between top and bottom at the onset of convection can be obtained. And, using this information and the properties of the working fluid there, the  critical dimensionless numbers, Marangoni and Rayleigh, 
can be calculated and compared with those found by Nield. 

As mentioned before, the previous approach present several drawbacks.  On one side, graphite tracers enable to observe the convection cells  but in addition to the perturb the flow, it does not provide information about the temperature field. On the other side,  thermocouples measure only the  temperature at a single point  and they also perturb the flow. Both problems can be solved  using a thermal camera. In such a case, graphite tracers are not needed to make the cells visible and the temperature field is directly observed. We describe the simplicity of the experiment in the next section.

\begin{figure}
    \centerline{\includegraphics[width=.78\columnwidth]{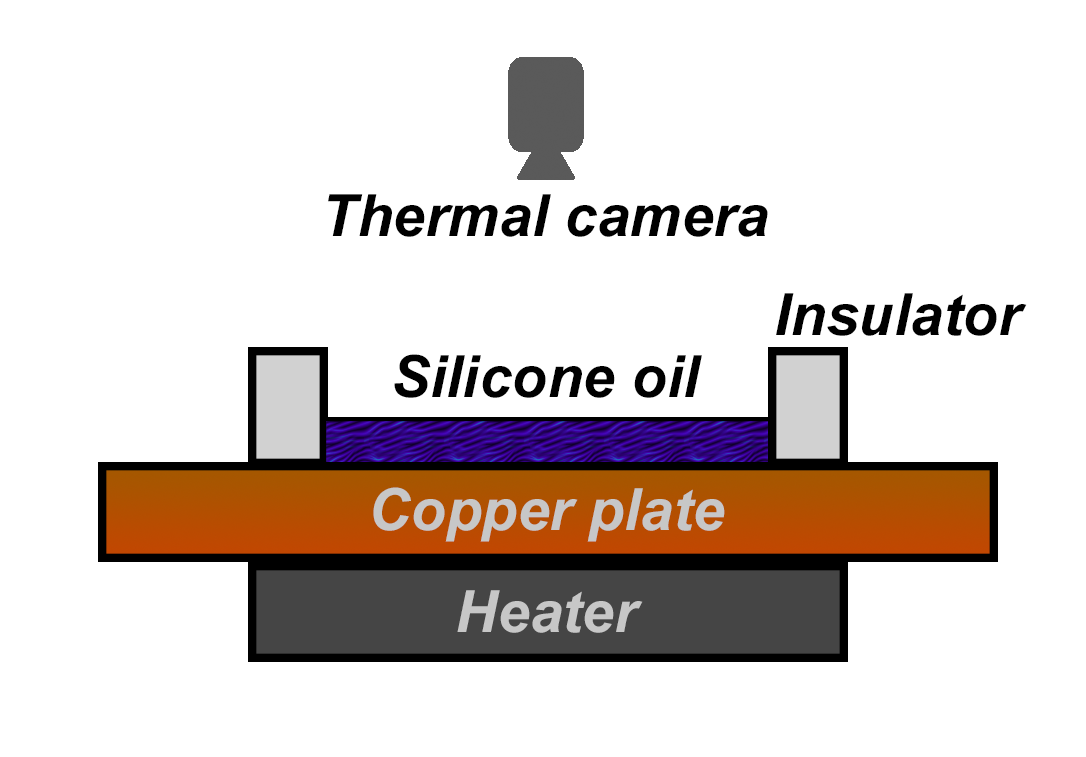}}
    \caption{Scheme of the experimental apparatus.}
\label{fig:Esquema}
\end{figure}

\section{The experiment}

The experimental apparatus, shown in Fig.~\ref{fig:Esquema}, consisted of a heater made with a modified electric kitchen heater (nominal power 500 W) covered with a bronze plate,  $2.4$ mm thick, to uniformly diffuse the temperature below a thin layer of silicone oil, and an insulator at the border. The inner diameter of the container was $D$ = $110$ mm.  
As a working ﬂuid we used silicone oil whose rheological properties were obtained from \cite{maroto2007introductory} where the same ﬂuid was
employed.

To register the thermal images, and then capture the temperature field, on the surface of the silicone oil above the apparatus we placed a thermal camera Flir One Pro iOS. The accuracy indicated by the manufacturer in the technical information is 3ºC or 5\% between the ambient and scene temperature in all the range. Nevertheless, in our experiments we compared with reference thermometers and in all the cases the differences were less than 3\%. The resolution was 160x120 pixels. In our setup 1 pixel corresponds approximately to 1 mm.
An important aspect is whether the thermal camera reliably records the temperature of the fluid layer. Experimental results for thin layers of 
silicone oil indicate that for wavelengths above  2600~nm  transmissivity is 
almost null \cite{siliconeoil} and, thus, the fluid absorbs almost all the 
radiation. In addition, under thermal equilibrium conditions, emissivity is 
nearly equal to the absorptivity, then,  the emissivity of the oil layer is 
close to 1.  As a consequence the thermal radiation registered by the camera 
corresponds to the temperature of the fluid layer. Nevertheless, prior to 
taking the measurements with the camera, the measured temperature was 
verified with reference thermometers. 

\begin{figure}
    \centerline{\includegraphics[width=.7\columnwidth]{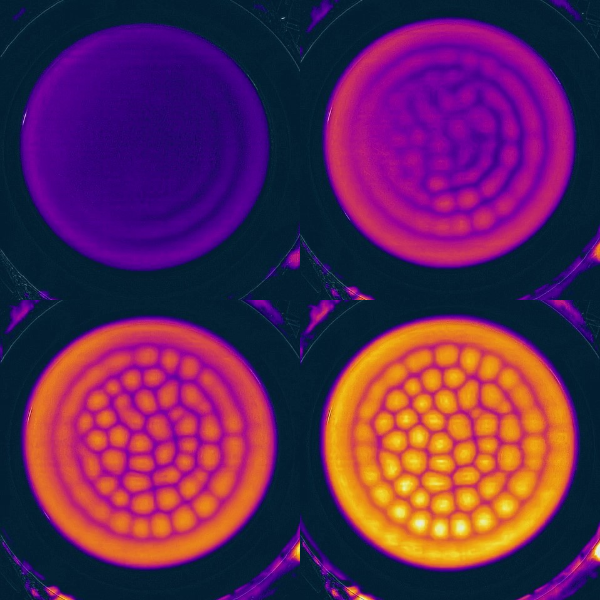}}
    \caption{Images obtained with the thermal camera while the fluid is warmed. The pattern began to be observed after
    a couple of  minutes.}
    \label{fig:Fotos}
\end{figure}

In our experiment we start with the fluid at room temperature and gradually heat it up with the kitchen heater. As soon as the temperature difference between the free surface and the bottom surface reaches a certain difference, the formation of convection cells can be seen  in the pictures shown in Fig.~\ref{fig:Fotos} \cite{cohen2004fluid}. 
These patterns began to be observed after about two minutes and stabilised in less than a minute. This interval depends on the thickness of the fluid but in all the cases remains in the order of a few minutes. In the figure  bright (dark) spots correspond to places where the temperature is higher (lower) so the fluid parcel rises (sinks). The experiment
can be repeated with different volumes of silicone oil and then
obtain temperature fields for layers of different thickness.

\section{Results and analysis}

\begin{figure}
\centerline{\includegraphics[width=.7\columnwidth]{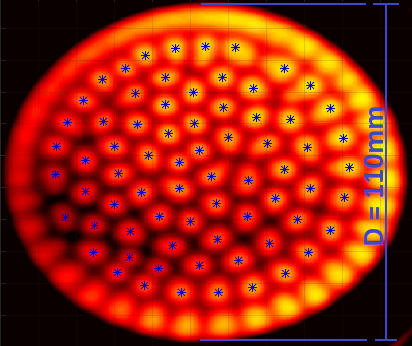}}
\caption{Spatial pattern obtained for fluid layer of thickness $d$ = 3.5 mm. The local maxima are indicated with  blue stars. }
\label{fig:Patron}
\end{figure}

As the fluid is heated from below, the temperature difference between the lower and upper surface increases. When this difference reaches a certain threshold, instability occurs and the typical hexagonal-shaped convection patterns are formed. During all this process, the temperature field is registered with  the thermal camera at 0.125 \textit{fps}. In Fig.~\ref{fig:Patron} we show the mean temperature of the peaks and the valleys in a small region. In a teaching laboratory context it would be important to highlight that these structures are not visible to the naked eye, so an important question is how to determine the  critical temperature from the digital images.

\begin{figure}
    \centerline{\includegraphics[width=.8\columnwidth]{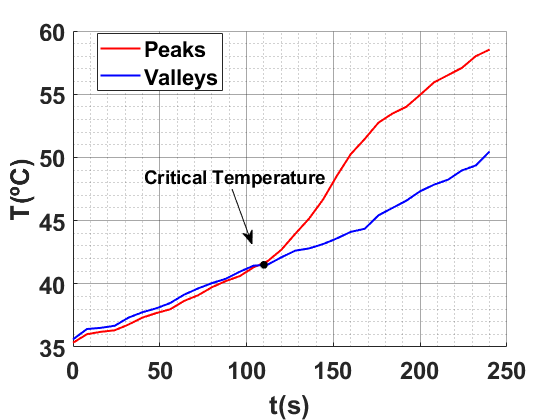}}
    \caption{Temperature of the peaks and the valleys as a function of time. 
    At the beginning of the experiment, both temperatures are equal, however, when a certain critical value is reached, there is a difference between the two. At this critical temperature the convection begins.}
    \label{fig:Critical_Temperature}
\end{figure}

Thermal image analysis serves two purposes. On the one hand, we can determine the critical temperature at the threshold of instability and on the other hand we determine the characteristic wavelength of the convection cells. In Fig.~\ref{fig:Critical_Temperature} we show the temporal evolution of temperature of the peaks and the valleys.
We observe that at the beginning of the experiment both temperatures are equal and they increase at the same rate. However when they reach a certain value they begin to diverge. This value corresponds to the critical temperature in which the base state loses stability and convection cells begin to form.

\begin{figure}
    \centerline{\includegraphics[width=.8\columnwidth]{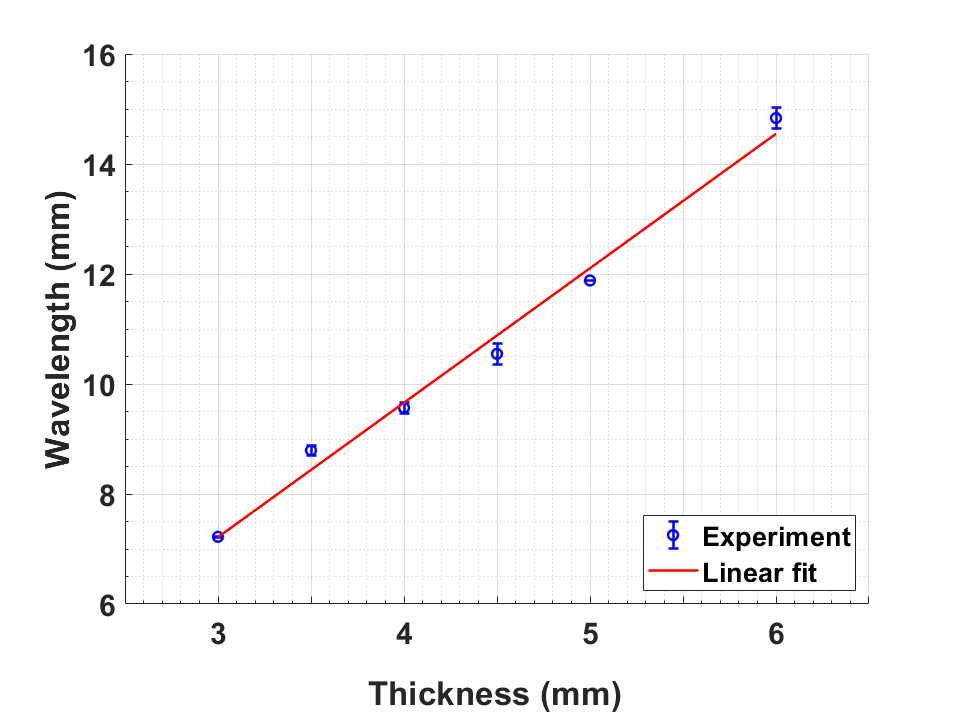}}
    \caption{Wavelength of the convection cells against the fluid layer thickness. The experimental data (blue) and a linear fit (red) are shown.}
    \label{fig:Ajuste}
\end{figure}

Image analysis is also used to determine the wavelength of the convection cells. Using  a Gaussian filter we found the pattern of the local maxima of the temperature field as shown in Fig.~\ref{fig:Patron}. 
The mean wavelength is obtained for each maximum averaging over the four nearest neighbours.  This procedure is repeated for different amounts of silicone oil which allow us to obtain results for different thicknes of the fluid layer. In Fig. \ref{fig:Ajuste}, we plot the wavelength  at the onset of instability against the layer thickness together with a linear fit of the experimental data. The correlation coefficient for a straight line is $0.993$ and the parameters of the equation $\lambda$ = $md$ + $n$ were $m$ = (2.44$\pm$0.13) and $n$ = (0.12$\pm$0.14) mm.

\begin{figure}
    \centerline{\includegraphics[width=.8\columnwidth]{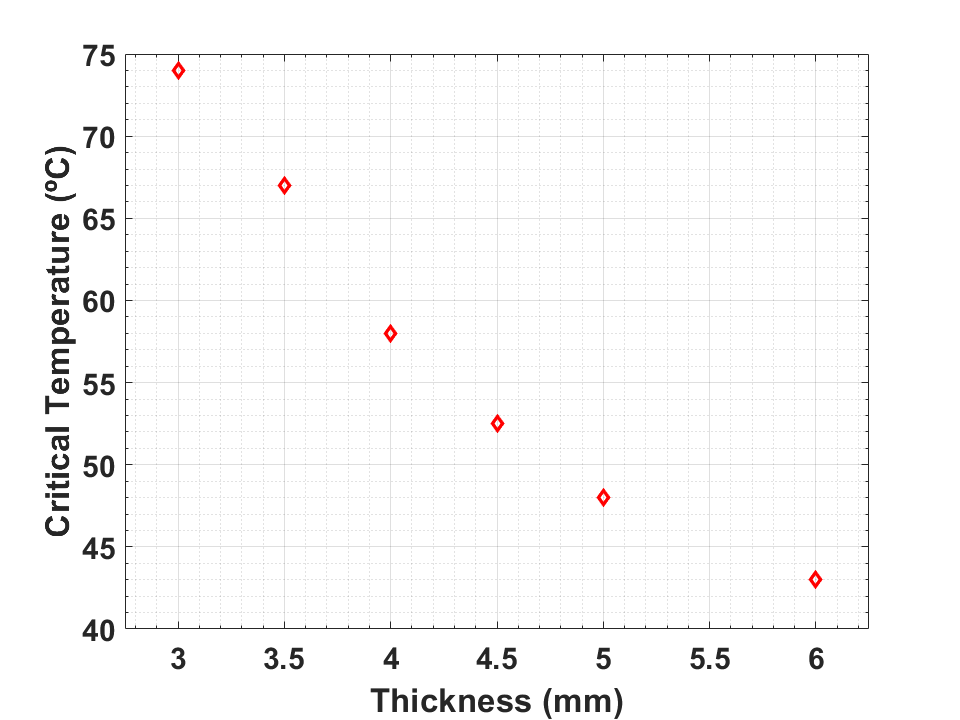}}
    \caption{Critical temperature for the onset of the convection cells, obtained using the procedure described in Fig.~\ref{fig:Critical_Temperature}, as a function of the thickness of the fluid layer. \label{fig:CriticaProf} }
\end{figure}

The critical temperature is plotted also as a function of the fluid layer thickness as shown in Fig.~\ref{fig:CriticaProf}.
In this thermo-capillary instability, the dependence between the critical temperature and the thickness of the fluid layer is difficult to model from a theoretical point of view. Indeed, both thermal effects, related to density variations with temperature, and capillary effects, related to surface tension variations with temperature, determine the critical 
parameter values of the instability  \cite{nield1964surface}. However, one aspect worth noting is that in the first case the instability depends on the Rayleigh number and in the second on the Marangoni number and in both cases the critical values show an inverse relationship between the temperature difference and the thickness of the fluid layer \cite{maroto2007introductory}. This dependence can be observed in this figure. A more exhaustive characterization of the phenomenon is beyond the scope of this paper.

\section{Conclusion}

In this work we propose a simple way to approach a traditional phenomenon of great importance such as thermocapillary instability. The proposed approach allows us to visualize the phenomenon without the need to add tracers while obtaining quantitative information that allows us to measure the critical temperature difference at which the well-known convection hexagons appear. By processing the images obtained it is possible to obtain the critical wavelength and in turn the relationship with the thickness of the fluid layer.

The results obtained for the wavelength and the critical temperature as
a function of the fluid layer thickness are simple to measure and according to those theoretically expected. We found a linear dependence of the wavelength with fluid thickness as predicted. The experimental set up is simpler than previous experiments and allows to experiment with different fluids in case it is desirable. The videos obtained with the thermal camera are striking examples of Bénard convection to use in lectures in fluid mechanics or nonlinear physics. 

This experiment was proposed to advanced undergraduate students (with previous knowledge of fluids, heat transfer and data processing) at our university and the results were positive. In all cases the students showed interest and enthusiasm in the proposal. The aspect that most attracted their attention was the fact that they were able to observe phenomena that are not visible to the naked eye and to obtain quantitative information from the thermal images.

\section*{Acknowledgements} \label{sec:ack}
The authors would like to thank PEDECIBA (MEC, UdelaR, Uruguay) and express their gratitude for the grant Fisica Nolineal (ID 722) Programa Grupos I+D CSIC 2018 (UdelaR, Uruguay).

\section{References}
\bibliography{iopart-num}


\end{document}